\documentstyle[aps]{revtex}
\begin{document}
\draft \twocolumn[\hsize\textwidth\columnwidth\hsize\csname
@twocolumnfalse\endcsname

\title{Black holes on the brane}

\author{Naresh Dadhich$^{1,a}$,
Roy Maartens$^{2,b}$, Philippos Papadopoulos$^{2,c}$
 and Vahid Rezania$^{2,3,d}$}

\address{~}

\address{$^1$Inter-University Centre for Astronomy and Astrophysics,
Ganeshkind, Pune~411~007, India}

\address{$^2$Relativity and Cosmology Group, School of Computer
Science and Mathematics, Portsmouth University,
Portsmouth~PO1~2EG, Britain}

\address{$^3$Institute for Advanced Studies in Basic Sciences,
Zanjan~45195, Iran}

\maketitle

\begin{abstract}

We consider exact solutions for static black holes localized on a
three-brane in five-dimensional gravity in the Randall-Sundrum
scenario. We show that the Reissner-N\"ordstrom metric is an exact
solution of the effective Einstein equations on the brane,
re-interpreted as a black hole {\em without electric charge,} but
with instead a {\em tidal `charge'} arising via gravitational
effects from the fifth dimension. The tidal correction to the
Schwarzschild potential is negative, which is impossible in
general relativity, and in this case only {\em one} horizon is
admitted, located outside the Schwarzschild horizon. The solution
satisfies a closed system of equations on the brane, and describes
the strong-gravity regime. Current observations do not strongly
constrain the tidal charge, and significant tidal corrections
could in principle arise in the strong-gravity regime and for
primordial black holes.

\end{abstract}

\pacs{04.50.+h, 04.70.-s \hfill{Phys. Lett. B, to appear}}
 \vskip 2pc
 ]

\section{Introduction}

Recent developments in string theory have shown that if matter
fields are localized on a 3-brane in $1+3+d$ dimensions, while
gravity can propagate in the $d$ extra dimensions, then the extra
dimensions can be large (see, e.g.,~\cite{add}). The extra
dimensions need not even be compact, as in the 5-dimensional
warped space models of Randall and Sundrum~\cite{rs}. (See
also~\cite{a} for earlier work.) In particular, they showed that
it is possible to localize gravity on a 3-brane when there is one
infinite extra dimension.

If matter on a 3-brane collapses under gravity, without rotating,
to form a black hole, then the metric on the brane-world should be
close to the Schwarzschild metric at astrophysical scales in order
to preserve the observationally tested predictions of general
relativity. Collapse to a black hole in the Randall-Sundrum
brane-world scenario was studied by Chamblin et al.~\cite{chr}
(see also~\cite{adm,gkr,ehm,cehs}). They gave a `black string'
solution which intersects the brane in a Schwarzschild solution.

Here we give an exact localized black hole solution, which
remarkably has the mathematical form of the Reissner-N\"ordstrom
solution, but {\em without electric charge} being present. Instead
the Reissner-N\"ordstrom-type correction to the Schwarzschild
potential can be thought of as a {\em `tidal charge',} arising
from the projection onto the brane of free gravitational field
effects in the bulk. These effects are transmitted via the bulk
Weyl tensor (see below). The Schwarzschild potential
$\Phi=-M/(M_{\rm p}^2r)$, where $M_{\rm p}$ is the effective
Planck mass on the brane, is modified to
\begin{equation}\label{0}
\Phi=-{M\over M_{\rm p}^2r}+{Q\over 2r^2}\,,
\end{equation}
where the constant $Q$ is a `tidal charge' parameter, which may be
positive or negative.

A geometric approach to the Randall-Sundrum scenario has been
developed by Shiromizu et al.~\cite{sms} (see also~\cite{bdul}),
and proves to be a useful starting point for formulating the
problem and seeing clear lines of approach. The field equations in
the bulk are (modifying the notation of~\cite{sms})
\begin{equation}
\widetilde{G}_{AB} =\widetilde{\kappa}^{2}\left[
 -\widetilde{\Lambda}\widetilde{g}_{AB}+
 \delta(\chi)\left( -\lambda
g_{AB}+T_{AB}\right)\right]\,, \label{1}
\end{equation}
where the tildes denote bulk quantities. The fundamental
5-dimensional Planck mass $\widetilde{M}_{\rm p}$ enters via
$\widetilde{\kappa}^2=8\pi/\widetilde{M}_{\rm p}^3$. The brane
tension is $\lambda$, and $\widetilde\Lambda$ is the bulk
cosmological constant. The brane is located at $\chi=0$ (so that
$x^4=\chi$ is a natural choice for the fifth dimension
coordinate), and $g_{AB}=\widetilde{g}_{AB}- n_An_B$ is the
induced metric on the brane, with $n_A$ the spacelike unit normal
to the brane. The brane energy-momentum tensor is $T_{AB}$, and
$T_{AB}n^B=0$. The brane is a fixed point of the $Z_2$ symmetry.

\section{Field equations on the brane}

The field equations induced on the brane arise from Eq.~(\ref{1}),
the Gauss-Codazzi equations and the matching conditions with
$Z_2$-symmetry, and they may be written as a modification of the
standard Einstein equations, with the new terms carrying bulk
effects onto the brane~\cite{sms}:
\begin{equation}
G_{\mu\nu}=-\Lambda g_{\mu\nu}+\kappa^2
T_{\mu\nu}+\widetilde{\kappa}^4S_{\mu\nu} - {\cal E}_{\mu\nu}\,,
\label{2}
\end{equation}
where $\kappa^2=8\pi/M_{\rm p}^2$. The energy scales are related
to each other and to the cosmological constants via
\begin{equation}
M_{\rm p}=\sqrt{{3\over4\pi}}\left({\widetilde{M}_{\rm p}^2\over
\sqrt{\lambda}}\right)\widetilde{M}_{\rm p}\,, ~~ \Lambda =
{4\pi\over \widetilde{M}_{\rm p}^3}\left[\widetilde{\Lambda}+
\left({4\pi\over 3\widetilde{M}_{\rm
p}^3}\right)\lambda^2\right]\,. \label{3}
\end{equation}
Typically, the fundamental Planck scale is much lower than the
effective scale in the brane-world: $\widetilde{M}_{\rm p}\ll
M_{\rm p} $. Local bulk effects on the matter are transmitted via
the `squared energy-momentum' tensor $S_{\mu\nu}$, but since we
will consider vacuum solutions, the precise form of $S_{\mu\nu}$
(see~\cite{sms}) will not be needed. In vacuum, $T_{\mu\nu}=0=
S_{\mu\nu}$, and we also choose the bulk cosmological constant to
satisfy $\widetilde{\Lambda}=-4\pi\lambda^2/3\widetilde{M}_{\rm
p}^3$, so that $\Lambda=0$ by Eq.~(\ref{3}). Then Eq.~(\ref{2})
reduces to
\begin{equation}\label{2'}
R_{\mu\nu}=-{\cal E}_{\mu\nu}\,,~~R_\mu{}^\mu=0={\cal
E}_\mu{}^\mu\,,
\end{equation}
where ${\cal E}_{\mu\nu}$ is the limit on the brane of the
projected bulk Weyl tensor~\cite{sms}:
\begin{equation}
{\cal E}_{AB}=\widetilde{C}_{ACBD}n^C n^D\,. \label{4}
\end{equation}
The Weyl symmetries ensure that this is symmetric and tracefree
(${\cal E}_{[AB]}=0={\cal E}_A{}^A$) and has no orthogonal
components (${\cal E}_{AB}n^B=0$, so that ${\cal E}_{AB} \to {\cal
E}_{\mu\nu} \delta_A{}^\mu \delta_B{}^\nu$ as $\chi\to 0$). It
carries the influence of nonlocal gravitational degrees of freedom
in the bulk onto the brane, including the tidal (or Coulomb) and
transverse traceless (gravitational wave) aspects of the free
gravitational field. (See \cite{m2} for a fuller discussion of
${\cal E}_{\mu\nu}$ in the general case.) On the brane, in the
vacuum case, this tensor satisfies the divergence
constraint~\cite{sms}
\begin{equation}
\nabla^\mu{\cal E}_{\mu\nu}=0\,, \label{5}
\end{equation}
where $\nabla_\mu$ is the brane covariant derivative. In view of
the Bianchi identities on the brane, this is an integrability
condition for the field equation $R_{\mu\nu}=-{\cal E}_{\mu\nu}$.
For static solutions, Eqs.~(\ref{2'}) and (\ref{5}) form {\em a
closed system of equations on the brane}~\cite{m2}.

A vacuum solution outside a mass localized on the brane must
satisfy equations (\ref{2'}) and (\ref{5}). This leads to a
prescription for mapping 4-dimensional general relativity
solutions to brane-world solutions in 5-dimensional gravity: {\em
a stationary general relativity solution with tracefree
energy-momentum tensor gives rise to a vacuum brane-world solution
in 5-dimensional gravity}. The 4-dimensional general relativity
energy-momentum tensor $T_{\mu\nu}$ (where $T_\mu{}^\mu=0$) is
formally identified with the bulk Weyl term on the brane via the
correspondence
\[
\kappa^2T_{\mu\nu}~\!\longleftrightarrow~\!-{\cal E}_{\mu\nu}\,.
\]
The general relativity conservation equations $\nabla^\nu
T_{\mu\nu}=0$ correspond to the constraint equation (\ref{5}) on
the brane. In particular, Einstein-Maxwell solutions in general
relativity will lead to vacuum brane-world solutions. This is the
observation that led us to the Reissner-N\"ordstrom-type solution.

\section{Solutions with tidal charge}

Algebraic symmetry properties imply that in general we can
decompose ${\cal E}_{\mu\nu}$ irreducibly with respect to a chosen
4-velocity field $u^\mu$ as~\cite{m2}
\begin{equation}
{\cal E}_{\mu\nu}=-\left({\widetilde{\kappa}\over\kappa}\right)^4
\left[{\cal U}\left(u_\mu u_\nu+{\textstyle {1\over3}}
h_{\mu\nu}\right)+{\cal P}_{\mu\nu}+2{\cal
Q}_{(\mu}u_{\nu)}\right]\,, \label{6}
\end{equation}
where $h_{\mu\nu}=g_{\mu\nu}+u_\mu u_\nu$ projects orthogonal to
$u^\mu$. Here
\[
{\cal U}=-\left({{\kappa}\over\widetilde\kappa}\right)^4 {\cal
E}_{\mu\nu}u^\mu u^\nu
\]
is an effective energy density on the brane arising from the free
gravitational field in the bulk---but note that this energy
density need not be positive. Indeed, as we argue below, ${\cal
U}<0$ is the natural case. The effective anisotropic stress from
the free gravitational field in the bulk is the spatially
tracefree and symmetric part, i.e.,
\[
{\cal P}_{\mu\nu}=-\left({{\kappa}\over\widetilde\kappa}\right)^4
\left[h_{(\mu}{}^\alpha h_{\nu)}{}^\beta-{\textstyle{1\over3}}
h_{\mu\nu}h^{\alpha\beta}\right]{\cal E}_{\alpha\beta}\,,
\]
where round brackets denote symmetrization. The effective energy
flux from the free gravitational field in the bulk is
\[
{\cal Q}_\mu=\left({{\kappa}\over\widetilde\kappa}\right)^4
h_\mu{}^\alpha{\cal E}_{\alpha\beta}u^\beta\,.
\]

In a static vacuum, with $u^\mu$ along the static Killing
direction, we have ${\cal Q}_\mu=0$, and the effective
conservation equation (\ref{5}) reduces to the single spatial
equation
\begin{equation}
{\textstyle{1\over3}}{\rm D}_{\mu}{\cal
U}+{\textstyle{4\over3}}{\cal U}A_\mu +{\rm D}^\nu{\cal
P}_{\mu\nu}+A^\nu{\cal P}_{\mu\nu}=0\,, \label{7}
\end{equation}
where ${\rm D}_\mu$ is the projection (orthogonal to $u^\mu$) of
the brane covariant derivative~\cite{m}, and $A_\mu= u^\nu
\nabla_\nu u_\mu$ is the 4-acceleration. Static spherical symmetry
means that
\[
A_\mu=A(r)r_\mu\,,~~{\cal P}_{\mu\nu}={\cal
P}(r)\left[r_{\mu}r_{\nu}-{\textstyle{1\over3}}h_{\mu\nu}\right]\,,
\]
for some functions $A(r)$ and ${\cal P}(r)$, where $r$ is the
areal distance and $r_\mu$ is a unit radial vector. The
Reissner-N\"ordstrom-type solution in Eq.~(\ref{0}) corresponds to
the solution
\begin{equation}\label{8}
{\cal U}=\left({\kappa\over\widetilde{\kappa}} \right)^4{Q\over
r^4}=-{\textstyle{1\over2}}{\cal P}
\end{equation}
of Eq.~(\ref{7}).

We can verify that the solution in Eqs.~(\ref{0}) and (\ref{8})
satisfies the field equations~(\ref{2'}), using natural
coordinates, for which the metric on the brane is
\begin{equation}\label{15}
ds^2=-A(r)dt^2+B(r)dr^2+r^2(d\theta^2+\sin^2\theta d\varphi^2)\,.
\end{equation}
Then it may be verified that
\begin{eqnarray}
&& A=B^{-1}=1+{\alpha\over r}+{\beta\over r^2}\,,\label{16}\\ &&
{\cal E}_t{}^t={\cal E}_r{}^r=-{\cal E}_\theta{}^\theta= -{\cal
E}_\varphi{}^\varphi={\beta\over r^4}\,,\label{17}
\end{eqnarray}
where $\alpha$ and $\beta$ are constants. Equations~(\ref{16}) and
(\ref{17}) satisfy all the field equations~(\ref{2'}), and hence
also the divergence equations~(\ref{5}). By Eqs.~(\ref{6}) and
(\ref{8}), we see that $\beta=Q$, and the far-field Newtonian
limit [see Eq.~(\ref{0})] shows that $\alpha=-2M/M_{\rm p}^2$.

In summary, we have shown that {\em an exact black hole solution
of the effective field equations on the brane is given by the
induced metric}
\begin{equation}\label{9}
-g_{tt}=(g_{rr})^{-1}=1-\left({2M\over M_{\rm p}^2}\right){1\over
r} +\left( {q\over\widetilde{M}_{\rm p}^2 }\right){1\over r^2}\,,
\end{equation}
{\em where $q=Q\widetilde{M}_{\rm p}^2$ is a dimensionless tidal
charge parameter.} The projected Weyl tensor, transmitting the
tidal charge stresses from the bulk to the brane, is
\begin{equation}\label{9'}
{\cal E}_{\mu\nu}=-\left( {q\over\widetilde{M}_{\rm p}^2
}\right){1\over r^4}\left[u_\mu u_\nu-2r_\mu r_\nu+h_{\mu\nu}
\right]\,.
\end{equation}

\section{Properties of the black hole}

The 4-dimensional horizon structure of the brane-world black hole
depends on the sign of $q$. For $q\geq0$, there is a direct
analogy to the Reissner-N\"ordstrom solution, with two horizons:
\begin{equation}\label{0'}
r_{\pm}={M\over M_{\rm p}^2}\left[1\pm\sqrt {1 -q{M_{\rm p}^4
\over M^2 \widetilde{M}_{\rm p}^2}}\,\right]\,.
\end{equation}
As in general relativity, both horizons lie inside the
Schwarzschild horizon: $0\leq r_-\leq r_+\leq r_{\rm s}=2M/ M_{\rm
p}^2\,$, and there is an upper limit on $q$:
\begin{equation}\label{11}
0\leq q\leq q_{\rm max}= \left({\widetilde{M}_{\rm p}\over M_{\rm
p}}\right)\left({M\over M_{\rm p}}\right)^2\,.
\end{equation}
The intriguing new possibility that $q<0$, which is impossible in
the general relativity Reissner-N\"ordstrom case, leads to only
{\em one horizon, lying outside the Schwarzschild horizon:}
\begin{equation}\label{0''}
r_{+}={M\over M_{\rm p}^2}\left[1+\sqrt {1 -q{M_{\rm p}^4 \over
M^2 \widetilde{M}_{\rm p}^2}}\,\right]> r_{\rm s}\,.
\end{equation}

In the $q<0$ case, the (single) horizon has a greater area than
its Schwarzschild counterpart, so that bulk effects act to {\em
increase the entropy and decrease the temperature of the black
hole.} In general relativity, the electric field in the
Reissner-N\"ordstrom solutions acts to {\em weaken} the
gravitational field, and the same is true for the brane-world
black hole with $q>0$. This can be seen clearly from
Eq.~(\ref{0}). By contrast, the $q<0$ case corresponds to the
opposite effect, i.e., bulk effects tend to {\em strengthen the
gravitational field.} By Eq.~(\ref{8}), we see that in this case,
the effective energy density ${\cal U}$ on the brane contributed
by the free gravitational field in the bulk is {\em negative.}
This is in accord with the (Newtonian) notion that the
gravitational field of an isolated mass has negative energy
density. Furthermore, it agrees with the perturbative analysis by
Sasaki et al.~\cite{ssm} and the nonperturbative analysis of
Maartens~\cite{m2}. The gravitational field generated by a source
on the brane tends to squeeze test matter in the 5th direction,
thus acting as an attractive field with a negative energy
contribution. The tidal acceleration measured by static observers
along the 5th direction $n^A$ is~\cite{m2}
\[
-\widetilde{R}_{ABCD}u^An^Bu^Cn^D=\left({\widetilde\kappa\over\kappa}
\right)^4{\cal
U}+{\textstyle{1\over6}}\widetilde{\kappa}^2\widetilde
{\Lambda}\,,
\]
where the right hand side follows from Eqs.~(\ref{1}) and
(\ref{4}). The negative bulk cosmological constant contributes to
acceleration towards the brane, reflecting its confining role on
the gravitational field. In order for ${\cal U}$ to reinforce
confinement, it must be negative. In other words, negative tidal
charge $q<0$ is the physically more natural case. (See also
\cite{v} for further discussion of negative energy density from
bulk effects.) Furthermore, $q<0$ ensures that the singularity is
spacelike, as in the Schwarzschild solution, whereas $q>0$ leads
to a timelike singularity, which amounts to a qualitative change
in the nature of the general relativistic Schwarzschild solution.

It is widely assumed that astrophysical black holes cannot exhibit
macroscopic electric charge due to the presence of neutralizing
plasma in their vicinity. Discussions of astrophysical black hole
phenomena are therefore commonly restricted to the Kerr family of
black holes. The solution presented here raises the possibility
that an effective Reissner-N\"ordstrom metric emerges by geometric
considerations and is hence not constrained by the above argument.
The tidal charge $q$ affects the geodesics and the gravitational
potential, so that indirect limits may be placed on it by
observations. Current observational limits on $|q|$ are rather
weak, since the correction term in Eq.~(\ref{0}) dies off rapidly
with increasing $r$, and astrophysical measurements (lensing and
perihelion precession) probe mostly (weak-field) solar scales.
These measurements require the correction term in the
gravitational potential to be much less than the Schwarzschild
term, so that
\begin{equation}\label{12}
|q| \ll 2\left({\widetilde{M}_{\rm p}\over M_{\rm p}}\right)^2
M_\odot R_\odot\,.
\end{equation}
This still allows for large values of $|q|$, which would modify
the spacetime geometry of a nonrotating black hole in the
strong-gravity regime~\cite{c}, with implications for example for
the last stable circular orbit for compact binaries. Although the
strong-gravity regime is not currently directly accessible to
observations, indirect limits could emerge from the way in which
tidal charge modifies general relativity strong-gravity effects.
This deserves further investigation.

Further indirect limits could arise from the effect of tidal
charge on primordial black holes, which also merits further
analysis. An intriguing possibility is that strong tidal effects
in the early universe could lead to black hole formation even in
the absence of gravitational collapse of matter. If such
matter-free tidal collapse is possible, then the endstate is a
metric with $M=0$ and $q<0$:
\begin{equation}\label{13}
-g_{tt}=(g_{rr})^{-1}=1+\left({q\over\widetilde{M}_{\rm p}^2}\right)
{1\over r^2}\,,
\end{equation}
and the horizon on the brane is given by
\begin{equation}\label{14}
r_{\rm h}={\sqrt{-q}\over\widetilde{M}_{\rm p}}\,.
\end{equation}

The tidal charge parameter arises from purely scalar
(Coulomb-like) effects of the free gravitational field in the
bulk, given the static spherical symmetry. In the original
Randall-Sundrum model, as well as in perturbative studies of it
(see, e.g.,~\cite{rs,gkr,cehs,ssm}), the bulk is assumed to be
exactly anti-de Sitter in the absence of any source on the brane.
The geometric approach of~\cite{sms}, which we have adopted here,
makes no assumptions about the bulk metric, other than that it
satisfies the 5-dimensional Einstein equations with cosmological
constant. Thus the bulk need not be conformally flat in the
absence of sources on the brane. This means that in general, {\em
the tidal charge parameter $q$ will be determined by both the
brane source, i.e., the mass $M$, and any Coulomb part of the bulk
Weyl tensor that survives when $M$ is set to zero.}

\section{Conclusion}

We have not investigated fully the effect of the brane-world black
hole on the bulk geometry, and in particular the nature of the
off-brane horizon structure. This has been done for solutions
which reduce to the Schwarzschild black hole on the
brane~\cite{chr}. In these solutions, the bulk metric is
\[
d\tilde{s}^2=-\left({6\over\widetilde{\kappa}^2\widetilde{\Lambda}}
\right){1\over z^2}\left[dz^2+g_{\mu\nu}(x^\alpha) dx^\mu
dx^\nu\right]\,,
\]
where $g_{\mu\nu}$ is the Schwarzschild metric. We have adopted a
different approach: instead of starting from an induced
Schwarzschild metric, we have solved the effective field equations
for the induced metric on the brane (which form a closed system,
since the metric is static), and found a generalization of the
Schwarzschild solution. It turns out that if $ds^2$ is given by
our solution, as in Eqs.~(\ref{15})--(\ref{9}), then
\[
d\tilde{s}^2=f(z)\left[dz^2+ds^2\right]
\]
cannot satisfy the bulk field equations for any $f(z)$ if $q\neq
0$. Finding an exact form of the bulk metric that is consistent
with our exact induced metric on the brane is more difficult than
the case where the induced metric is Schwarzschild.

Perturbative studies, which start from an exactly anti-de Sitter
background, show that the first weak-field correction of the
Newtonian potential on the brane is proportional to $1/r^3$
(see~\cite{rs,gkr,cehs,ssm}). Our solution has by contrast a
$1/r^2$ correction, so that it is incompatible with the
long-distance limit on the brane. (Such a correction can also
arise in thick-brane models~\cite{cehs}.) However, in the {\em
short-distance} limit on the brane, the lowest order correction to
the potential is proportional to $1/r^2$~\cite{s}. This term will
dominate the $1/r$ term, which reflects the fact that gravity
becomes effectively 5-dimensional at high energies. Thus our
solution should describe well the strong-gravity regime on the
brane. In the short-distance limit, the perturbative analysis
shows that~\cite{s}
\[
q=-{M\over\widetilde{M}_{\rm p}}\,,
\]
so that the tidal charge is negative, as we argued above, and is
determined by the black hole mass (which is to be expected if the
background bulk Weyl tensor vanishes).

In order to pursue our nonperturbative analysis, we need to look
at the off-brane equations for the curvature (see~\cite{ssm} for
the general form of these equations). The induced field equations
on the brane given in Eq.~(\ref{2'}) are supplemented by off-brane
equations which determine
\[
{\cal L}_{\bf n}{\cal E}_{AB}\,,~ {\cal L}_{\bf n}{\cal
B}_{ABC}\,,~ {\cal L}_{\bf n} R_{ABCD}\,,~
\]
where ${\cal L}_{\bf n}$ is the Lie derivative along $n^A$,
$R_{ABCD}$ is the 4-dimensional Riemann tensor, and
\[
{\cal B}_{ABC}=g_A{}^Dg_B{}^E\widetilde{C}_{DECF}n^F\,,
\]
with ${\cal B}_{\mu\nu\sigma}=0$ on the brane. These equations
together with Eq.~(\ref{2'}) form a closed system~\cite{ssm}.

In conclusion, we have shown how a Reissner-N\"ordstrom-type
metric satisfies the effective field equations on the brane in
Randall-Sundrum-type gravity, which form a closed system because
of staticity. There is no electric charge present, but instead a
tidal charge, arising from the imprint of the free gravitational
field in the bulk on the brane. The tidal charge correction to the
Schwarzschild potential is negative, and the solution describes
the strong-gravity regime on the brane. Negative $q$ leads to an
horizon on the brane that is outside the Schwarzschild horizon,
corresponding to lower temperature and greater entropy. Current
observations place only weak limits on the tidal charge, and in
principle significant tidal modifications could arise in the
strong-field regime or in the early-universe case of primordial
black holes. Further investigation is in progress to probe these
modifications and any indirect limits that they may impose on the
tidal charge, as well as to find the off-brane behaviour of the
horizon.

\[ \]{\bf Acknowledgements:} VR was supported by a Royal Society grant
while at Portsmouth, and thanks the Relativity and Cosmology Group
for hospitality. RM thanks IUCAA, Pune for hospitality during a
visit, which was partially supported by the Royal Society. We
thank Bruce Bassett, Marco Bruni, Roberto Casadio, Roberto
Emparan, David Langlois, Jose Senovilla, Carlo Ungarelli and David
Wands for useful discussions and comments, and especially Tetsuya
Shiromizu.

\end{document}